\documentstyle[aps,prl,multicol,epsf]{revtex}

\def \be{\begin{displaymath}}
\def \ee{\end{displaymath}}              

\def \ben{ \begin{equation} }
\def \een{ \end{equation}   }            

\def \bea{\begin{eqnarray*}}             
\def \eea{\end{eqnarray*}}

\def \bean{\begin{eqnarray}}             
\def \eean{\end{eqnarray}}

\def \nn{\nonumber}

\def \Knu#1{\mbox{K}_{#1}}
\def \Knull{\Knu{0}}

\def \Ref#1{(\ref{#1})}

\def \inv{ ^{-1} }
\def \dag{^\dagger}

\def \invb#1 { \frac{1}{#1} }

\def \av#1{ {\left\langle #1 \right\rangle} }
\def \dx{\partial_x}
\def \dt{\partial_t}
\def \d{\mbox{d}}

\def \fr#1#2{ \frac{#1}{#2} }

\def \gbar{ \bar g }
\def \dbar{ \bar d }
\def \lambar{ \bar \lambda }
\def \half    { \frac{1}{2} }

\def \lapprox{ \stackrel{<}{\sim} }
\def \gapprox{ \stackrel{>}{\sim} }

\begin{document}

\title{ Coulomb Drag between Quantum Wires}
 \author{ Rochus Klesse\footnote{ present address: Universit\"at zu K\"oln,
     Institut f\"ur Theoretische Physik,
     Z\"ulpicher Str. 77, D-50937 K\"oln, Germany.} and Ady Stern}

 \address{Department of Condensed Matter Physics, The Weizmann
 Institute of Science, Rehovot 76100, Israel}

\date{\today}

\maketitle

\begin{abstract}
We study Coulomb drag in a pair of parallel
one-dimensional electron systems within the framework of 
the Tomanaga-Luttinger model.
We find that Coulomb coupling has a much stronger effect on one dimensional
wires than on two-dimensional layers: At zero temperature the
trans-resistivity {\em diverges}, due to the formation of  
locked charge density waves.
At temperature well above a cross-over temperature $T^*$ 
the trans-resistivity follows a power law $\rho \propto T^x$, where
the interaction-strength dependent exponent $x$
is determined by the Luttinger Liquid parameter $K_{c-}$ of the
relative charge mode. 
At temperature below $T^*$ relative charge displacements are enabled by
solitonic excitations, reflected by an exponential
temperature dependence.
 The cross-over temperature $T^*$
depends sensitively on the wire width, inter-wire distance, Fermi
wavelength and the effective Bohr radius. 
For wire distances $\bar{d} \gg k_F^{-1}$ it is exponentially suppressed
with $T/E_F \sim \exp[ - \bar{d} k_F / (1-K_{c-}) ]$.
The behavior changes drastically if each of the two  wires develop spin 
gaps. In this case we find that the trans-resistivity {\em vanishes} at
zero temperature. We discuss our results in view of possible
experimental realizations in GaAs-AlGaAs semiconductor structures.

\end{abstract}
\pacs{PACS: 71.10.Pm, 73.23.Ad}
\begin{multicols}{2}
\section{Introduction}

Measurements of Coulomb drag trans-resistivity between two coupled low
dimensional electronic systems are a powerful probe of scattering and
correlations between electrons \cite{2Ddrag}. In a measurement of the
trans-resistivity $\rho_D$ a current $I_1$ is driven in one (the
"active") of the systems, while no current is allowed to flow in the
other system (the "passive" system). The Coulomb interaction between
electrons in the two systems transfers momentum from the active system
to the passive one, where a voltage drop $V_2$ develops. The ratio
$-V_2/I_1$ is the trans-resistance, which is related to the
trans-resistivity by a geometric factor.

In weakly coupled two dimensional systems, at least at zero magnetic
field, the trans-resistivity is usually proportional to the
electron--electron momentum relaxation time, and is therefore
proportional to $T^2$, with $T$ being the temperature. As explained by
Fermi liquid theory, the $T^2$ behavior holds also in the presence of
electron--electron interaction within each of the two coupled systems.

In one dimensional systems, which are presently realized by organic
quasi-1D-metals, carbon
nanotubes, edge states of quantum hall liquids, and 1D semiconducting
structures, electron--electron interaction is believed to invalidate
the Fermi liquid picture, and generate a different state, described
approximately by the Tomanaga--Luttinger (TL) theory
\cite{tomanaga__luttinger,haldane} (for reviews see e.g.
\cite{voit,emery,solyom,schulz}). Since 
electronic correlations in this state are stronger than in a Fermi
liquid, it is interesting to examine the Coulomb drag trans-resistivity
between two such systems.  

In this paper we study theoretically Coulomb drag between two identical parallel one 
dimensional wires at close proximity. For perfectly clean wires, as
assumed here, the current flowing in the active wire generates
voltages on the two wires, which, due to Galilean invariance, are
equal in magnitude and opposite in sign. The trans-resistivity
$\rho_D\equiv -\frac{V_2}{I_1L}$ (with $L$ the length of each wire) is
then also the intrinsic resistivity (not including the contact resistance)
 of the active wire. Thus, we occasionally
refer to $\rho_D$ as the ``resistivity''. Note, however, that this
resistivity does not influence a symmetric flow of current in the two
wires.

Drag between 1D electron systems was considered earlier by several 
authors. Hu and Flensberg \cite{hu_flensberg} 
and more recently Raichev and Vasilopoulos \cite{raichev} 
investigated the problem in the absence of electron correlations
(apart from screening effects) 
within Fermi liquid theory.
Tanatar \cite{tanatar} studied the
same problem in the presence of disorder, and Coulomb drag 
of Luttinger liquids with a point-like interaction region was
considered by Flensberg \cite{flensberg}, and Komnik and Egger 
\cite{komnik}.
In a recent work by Nazarov and Averin \cite{nazarov_averin} 
1D systems of spin-less electrons are treated as independent Luttinger liquids with
coupling limited to inter-wire backscattering ($\Delta k \approx 2k_F$,
where $k_F$ is the Fermi wavevector in the two wires).
The present
work treats both intra- and inter-wire electron-electron interaction
on equal footing. We find that although drag takes place primarily
through $2k_F$ scattering,  
the small momentum component of the inter-wire interaction and 
spin-density interactions affect it in a crucial way. 
The problem under consideration is also closely related to 
the  problem of a  coupled double (or $N-$)chain
\cite{lee_rice_klemm,double_chain}. In case of the spin-full
problem results of Lee et al.
\cite{lee_rice_klemm} are useful.

The paper is organized as follows: In section \ref{results}
we define the problem and present the main results and the physical
picture. Section \ref{spinless} deals with two wires 
of spin-less electrons (throughout the paper we use ``wire'' as 
synonym for ``1D electron system''). After formulating this problem
in subsection \ref{sl-ham}, we analyze it by means of a
renormalization group in \ref{rg}. We then discuss the high
  temperature regime (\ref{high_temperature}), the
  crossover temperature (\ref{xovert}), and the low
  temperature regime (\ref{low_temperature}). In this
  analysis we employ the method developed recently by Nazarov and
  Averin \cite{nazarov_averin}, and earlier results on impurity pinned
  charge density waves (CDWs) \cite{rice,maki}. In section \ref{spinfull}
we address a double-wire system of spin unpolarized
electrons.  We write the Hamiltonian and renormalization group
equation in subsection \ref{sfham}, analyze the fixed points in
\ref{sffp}, deal with weak interactions in \ref{sfsmallu}, and extend
the discussion beyond that limit in 
subsection \ref{sfallu}. We then  discuss the high and low temperature
limits of the drag in this case in subsection \ref{sfhlt}. In section
\ref{experiments}  
we estimate experimental values of the relevant parameters for
semiconducting wires \cite{tarucha,yacobi}.  
section \ref{comparison} concludes with a summary. 
Some technical details are put into the appendices. In particular, in
appendix \ref{equivalence} we examine the relation between the
Nazarov-Averin method and earlier weak coupling calculations of
Coulomb drag \cite{zheng}.

\section{Review of the main results}\label{results}

We consider two identical wires of diameter $d$, separated by a
distance $\dbar$. We denote by $k_F$ the Fermi wave vector in each
wire, by $v_F$ the Fermi velocity in each wire. The strength of the
Coulomb interaction is characterized by $r_s = r/a_B$, with $r$ 
the mean (intra-wire) electron distance and $a_B$ the effective 
Bohr radius. 
The length of the two wires is $L$.  

We first consider two wires of spin-less electrons. Experimentally, this system may be realized by applying a magnetic field parallel to the wires, which would polarize the electrons' spins without affecting their orbital motion. This system is closely related to a single Luttinger liquid with a spin-degree of freedom, when the
two spin-projections are identified with electrons in the two wires.
Therefore, results obtained previously on the effect of backscattering in such 
systems \cite{luther_emery,chui_lee} can be used.

We find that, for infinitely long wires,  Coulomb coupling
always leads to a {\em diverging} resistivity
$\rho_D$ as temperature goes to zero. The physical picture behind this
effect is that at sufficiently low temperature the electrons in both wires form two inter-locked
CDWs. Then a relative charge
displacement can be created only by overcoming a potential barrier. At zero temperature this cannot be done by an infinitesimal electric field, and leads to a non-linear trans-resistance. At finite $T$, below a crossover temperature $T^*$ (discussed below), the trans-resistance satisfies,  
\be
\rho(T) \sim \rho_{0,T} \exp(E_s/T)
\ee
with $E_s\sim T^*$ defined below. 

For short wires a
qualitative different behavior appears. Here, from time to time
the CDW in the active wire slips as a whole 
relative to the CDW in the other wire. These
instantaneous slips are a result  of either thermal fluctuations or 
tunneling events. 
The latter leads to a non-diverging resistance at zero temperature,
which is exponential in $L$.

At temperatures well above $T^*$ the previous picture of inter-locked
CDWs is no longer valid. In this case it is
more appropriate to think of independent electrons in the active wire,
which suffer from backscattering at the $2k_F$ component of the potential generated by density fluctuations in the passive wire. 
A perturbative calculation yields in this case a resistivity
\ben
 \rho(T) = \rho_0 \lambda^2 \left( \fr{T}{E_0} \right)^{\scriptsize x},
\quad x = 4K_{c-}-3,
\label{hight}
\een 
exhibiting a characteristic power law dependence on $T$. 
The coefficient $\rho_0$ is of order $hk_F/e^2$,
$\lambda$ denotes the 
dimensionless inter-wire backscattering potential, and $E_0$ is of
order the Fermi energy. The  parameter $K_{c-}$ is
the TL parameter of the {\em relative} charge density sector ($c-$).
It is determined by the {\em difference} of the small
momentum intra- and inter-wire couplings, and not by the intra-wire
small momentum coupling, as assumed in \cite{nazarov_averin}. 
With vanishing small-momentum interaction, $K_{c-}$ approaches unity, 
and $\rho(T)$ takes the  linear temperature 
dependence  of the drag resistance of independent 
1D electrons  \cite{hu_flensberg}. In the presence of Coulomb interactions $K_{c-}$ may be either larger or smaller than $1$, depending on the inter-wire distance. 

The crossover temperature $T^*$ is a complicated function of four
length scales: the wire separation $\dbar$, wire width $d$,
effective Bohr radius $a_B$, and the mean (intra-wire) electron
distance $r=\pi/k_F$.  The first, $\dbar$, controls $\lambar$, the
strength of the $2k_F$ component of the inter-wire Coulomb
interaction. For widely separated wires $\dbar\gg r$, this component
is exponentially small, $\lambar\propto \exp{-2 k_F \dbar}$ and
consequently,  
\be
 T^* \sim E_0 \exp\left(- \fr{k_F \dbar}{1-K_{c-}} \right).
\ee
As $\dbar/r\rightarrow\infty$, then, $T^*\rightarrow 0$ and the
trans-resistivity follows Eq. (\ref{hight}) in all practically
relevant temperatures and length scales. 
The general trends are shown in table \ref{table-one}. 
The maximum values of $T^*$ that can be expected in realistic
experimental set-ups are of order $T^* \sim 0.01 \times E_0$.
In case of small wire separation $\dbar\ll r$
the crossover temperature $T^*$ is exponentially suppressed according
to 
\be
T^* \sim E_0 \exp\left(-\fr{\pi^3}{r_s}\fr{c(k_F)}{k_F \dbar}\right),
\ee
where $c(k_F)$ is of order one and only logarithmically depends on $k_F$.

For the spin-full case the results are similar, as long as the spin
sectors are not unstable towards a formation of an energy gap in their
spectrum (spin gap). The resistivity diverges at zero temperature, and
scales with temperature with an exponent
\be
 x = 2K_{c-} -1
\ee
in the high temperature regime. A comparison with the previous 
exponent reveals that fluctuations in the neutral spin sector moderate
the effect of the charged modes.

The behavior changes drastically if the single wires develop spin 
gaps. In this case we find that the trans-resistivity {\em vanishes} at
zero temperature.

\section{Spin-less double wire}\label{spinless}
\subsection{Notation}
We use the following notations:
\begin{description}
\item[$a_{rw}\dag(k), a_{rw}(k)$] : creation and annihilation operator
  of a right ($r=+$) or left ($r=-$) moving fermion of momentum $k$.
  The second index refers to the active ($w=+$) and passive ($w=-$) wire. 
\item[$\psi_{rw}^{(\dagger)}(x) = L^{-1/2} \sum_k e^{ikx}
  a_{rw}^{(\dagger)}(k)$   ]  : fermionic operators in real space
  representation.
\item[$\rho_{rw}$] : density of right/left moving fermions in wire
  $w$.
\item[$n_w = \rho_{-,w} + \rho_{+,w}$] : charge density of wire $w$.
\item[$n_{c\pm} = n_- + n_+$] : absolute (or symmetric, $c+$) and relative
  (or antisymmetric, $c-$) charge density of the double wire system.
\end{description}
In general, the indices $c+$ and $c-$ refer to quantities of the
absolute and relative charge mode, respectively. 
We also use this convention for bosonic field, introduced in the next
subsection.  
  
The notation we use for the coupling constants follows that of Voit's
review \cite{voit}. 
We use $g_{i}$ to denote intra-wire couplings, and
$\gbar_i$ to denote inter-wire couplings. The subscript $i=1$ denotes
$2k_F$ scattering, $i=2,4$ denote small momentum scattering.  

\subsection{The Hamiltonian in fermionic and bosonic
  representation}\label{sl-ham} 
In this section we consider two one dimensional wires of spin polarized
electrons with equal densities. If the wire index is viewed as a
$z$-component of an "iso-spin", the Hamiltonian of the problem is that
of an iso-spin-$\half$ 1D system (but with interactions that are not
$SU(2)$-symmetric),  and results obtained previously on the
effect of backscattering in such systems \cite{luther_emery,chui_lee}
can be used.
 The kinetic energy part of the Hamiltonian states
\cite{voit}, 
\bean
H_0 = v_F \sum_{rwk} rk a_{rw}\dag(k) a_{rw}(k) \label{h_0a}
= \fr{\pi v_F}{L} \sum_{qrw} \rho_{rw}(q)\rho_{rw}(-q) \label{h_0b}.
\eean 
The small-momentum transfer or forward scattering
part of the electron-electron interaction is given by \cite{voit}
\bea
H_f = &\quad &
    \fr{1}{L} \sum_{ww'q}
    [ \delta_{w,w'} g_2 + \delta_{w,-w'} \gbar_2 ]
     \rho_{+w}(q)\rho_{-w'}(-q) \\
 &+&
    \fr{1}{2L} \sum_{rww'q}
    [ \delta_{w,w'} g_4 + \delta_{w,-w'} \gbar_4 ]
     \rho_{rw}(q)\rho_{rw'}(-q)
\eea
and backscattering processes are described by \cite{voit}
\bea
H_b &=& \sum_{ww'} \int \! dx\: \psi_{+w}\dag(x) \psi_{-w'}\dag(x)
\psi_{+w'}(x) \psi_{-w}(x) \times \\
& & \qquad\qquad[ \delta_{w,w'}g_1 + \delta_{w,-w'}\gbar_1 ].
\eea

It is convenient to switch to a standard boson representation
by introducing bosonic fields $\phi_w(x) = -i\pi L\inv\sum_q 
q\inv e^{-iqx-\alpha|q|/2}( \rho_{+w}(q) + \rho_{-w}(q) )$
with their conjugates
$\Pi_w(x) = L\inv \sum_q 
e^{-iqx-\alpha|q|/2}( \rho_{+w}(q) - \rho_{-w}(q) )$ \cite{schulz}.
Throughout the paper we  interpret the length $\alpha$ as 
the inverse Fermi wavevector $2\pi/k_F$.
Physically, the field $\phi_w(x)$ denotes the displacement
of electrons in wire $w$, normalized in such a way that 
density fluctuations $\delta n_w(x)$ and current $I_w(x)$ are given 
by $\dx \phi_w(x) = - \pi \delta 
n_w(x)$ and $\dt \phi_w(x) = \pi I_w(x)$. The relation
to the fermions $\psi_{rw}$ is established by the Luther-Peschel
transformation formula, $\psi_{rw}(x) = (2\pi \alpha)^{-1/2} 
\exp[ir(k_F x - \phi_w(x)) + \theta_w(x)]$, where $\theta(x) =
\pi \int_{x_0}^x \!dx' \Pi_w(x')$ \cite{schulz}.

The total Hamiltonian
$H = H_0+ H_f + H_b$ separates into two decoupled parts, one describing  absolute (symmetric) current and density, 
and one describing relative (antisymmetric) current and density. The decoupling is obtained by means of the transformation,  $\phi_{c\pm}=2^{-1/2}(\phi_+ \pm \phi_-)$,
$\Pi_{c\pm}=2^{-1/2}(\Pi_+ \pm \Pi_-)$
\cite{luther_emery}. 
In bosonic representation the two parts are\cite{voit}
\bean
H_{c+} &=& \fr{u_{c+}}{2\pi} \int dx 
 \left(
    K_{c+} \pi^2 \Pi^2_{c+} +
   \fr{1}{K_{c+}}(\dx \phi_{c+})^2 
 \right),\label{H_absolut}             \\
H_{c-} &=& \fr{u_{c-}}{2\pi} \int dx
 \left(
    K_{c-} \pi^2 \Pi^2_{c-} + 
   \fr{1}{K_{c-}}(\dx \phi_{c-})^2 
 \right) \\
& & \quad\quad\quad + \fr{2 \gbar_1}{(2\pi \alpha)^2 } 
   \int dx \cos( \sqrt{8} \phi_{c-} ) \label{H_relative}, \\
& & K_{c\pm} = \sqrt{\frac{ 1 + U_{c\pm} }{ 1 -  U_{c\pm} }}, \label{kcpm}\\ 
& & \quad U_{c+} = \fr{1}{2\pi v_{c+}}( - g_2 - \gbar_2 + g_1), \\
& & \quad U_{c-} = \fr{1}{2\pi v_{c-}}( - g_2 + \gbar_2 + g_1), \label{ucm} \\
& & \quad\quad  v_{c\pm} = v_F + ( g_4 \pm \gbar_4)/2\pi, \\
& & u_{c\pm} = v_{c\pm} (1- U^2_{c\pm})^{1/2}.
\eean
(The signs in the definition of the small momentum couplings
$U_{c\pm}$ are chosen according to the conventions in \cite{chui_lee}.)
The fields $\phi_{c\pm}$ describe fluctuations in the absolute
and relative density via $\partial_x \phi_{c\pm} = - 2^{-\half}\pi
\delta n_{c\pm}$. Accordingly, the relation to currents are $\dt
\phi_{c\pm} =  2^{-\half}\pi I_\pm $.

The current in a drag experiment is a superposition of a symmetric and
an anti-symmetric current.  A symmetric current flows without
resistivity, due to Galilean invariance. Thus, the resistance results
from the antisymmetric part only, and is  
determined by the relative charge sector $H_{c-}$ only. Formally this
is manifested in the invariance of $H_{c+}$ to spatially homogeneous 
charge displacements $\phi_{c\pm}(x) \to \phi_{c\pm}(x) +
\varphi_{c\pm}$, an invariance which is absent in the backscattering potential ($\propto
\cos\sqrt{8} \phi_{c-}$) in $H_{c-}$. 
 Consequently, we confine
ourselves  to the sine-Gordon type Hamiltonian $H_{c-}$.

The Hamiltonian $H_{c-}$ has two parameters, $K_{c-}$ (which may be
expressed in terms of $U_{c-}$, see Eq. (\ref{kcpm})) and
$\gbar_1$. Our results are all independent of the sign of $\gbar_1$,
which we take below to be positive.
For a single wire with electrons of two spin directions (namely, for $SU(2)$
symmetric interaction, $g_2=\gbar_2$) the sign of $U_{c-}$ is
determined by the sign of $g_1$, i.e., by whether the interaction is
repulsive or attractive.
Here this only holds for two wires which are very close to one
another. For larger inter-wire distances $U_{c-}$ can become negative
also for repulsive interaction, in particular Coulomb interaction: 
Then, $\gbar_2-g_2=-\fr{2e^2}{\epsilon} \ln{\dbar/d}$, while the
parameter $g_1$ is independent of $\dbar$. 
Thus, for large inter-wire distance $\dbar$, the parameter $U_{c-}<0$.

Below we confine ourselves to repulsive interaction, and
discuss the case of wires at close proximity and that of well
separated wires.

\subsection{Renormalization group analysis}\label{rg}

In this subsection we analyze the backscattering term of the
Sine-Gordon Hamiltonian by means of a renormalization group (RG)
analysis, and show that if the bare interactions are weak, and the
electron-electron interaction potential decays with distance, the drag
resistivity diverges at zero temperature.  

 Let us first recall the main elements of an
RG treatment for $H_{c-}$ (see e.g. \cite{voit}).
For small backscattering couplings, 
$\lambar \equiv \gbar_1/2\pi u_{c-} \ll 1$, the RG equations
are of the Kosterlitz--Thouless type (here we denote $K\equiv K_{c-}$) 
\ben\label{flow} 
\fr{ d \lambar}{\d x} = (2 - 2K) \lambar, \quad 
\fr{ d K }{\d x} = - 2 \lambar^2 K^2,
\een
where the parameter $ x = \ln l/\alpha $ is the logarithm of 
the renormalized momentum cut-off $l\inv$. The RG procedure
starts with the bare couplings $\lambar_0,K_0$ at an initial momentum
cut-off $l_0\inv$ of order of $\alpha^{-1}$ and ends with renormalized 
couplings at a final cut-off $l_1\inv=\max\{ L^{-1},T/u_{c-} \}$.
One method (due to Jose et al. \cite{jose})
to derive the RG equations is to expand the scale invariant
correlation function
$\av{e^{2i\phi_{c-}(x_1,\tau_1)}e^{-2i\phi_{c-}(x_2,\tau_2)}}$ in 
powers of the coupling $\lambar$ and to integrate out the large
momentum degrees of freedom. After re-exponentiating the result
one can then read off the RG equations (for details see 
\cite{jose} and also \cite{giamarchi_schulz}). Within each $RG$ step 
only the backscattering interaction $\gbar_1$ is treated
perturbatively, whereas the small momentum interaction parameters
$g_2-g_1,g_4,\gbar_2,\gbar_4$, which determine the parameter 
$K$, are treated exactly. 
A different method, which
leads to the same RG equations (\ref{flow}) but may provide 
additional insights, is by means of mapping the Sine--Gordon
Hamiltonian onto the two-dimensional Coulomb gas problem
\cite{chui_lee}.  

\begin{figure}[htb] 
 \begin{center}
         \epsfxsize 8cm
         \epsffile{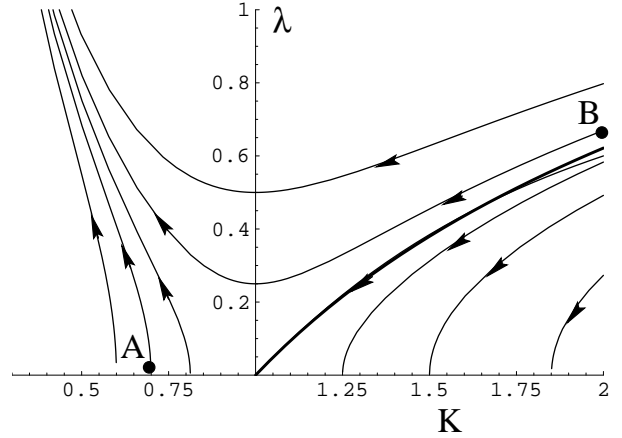}
 \end{center}
 \caption{ The RG-flow of a double wire system of spin-less
         electrons. Point A corresponds  to the bare couplings of a
 system  with $ \bar{d}  \gg k_F\inv $, 
 point B to wires with narrow spacing $ \bar{d} \ll k_F\inv $.
 }
 \label{fig-flow}
 \end{figure}

The integral curves $\lambar(K)$ of the RG-flow \Ref{flow} shown in 
Fig. \Ref{fig-flow} obey the differential equation
\be
\fr{\d \lambar}{\d K} = \fr{K-1}{K^2 \lambar},
\ee
and are of the form
\be
\lambar(K) = \sqrt{2} \left( \fr{1}{K} + \ln(K/K_0) -\fr{1}{K_0} +
\fr{\lambar_0^2}{2} \right)^{1/2}.
\ee

There are two types of stable fixed points to Eqs. (\ref{flow}). Fixed points of the  first type are characterized by $\lambar=0,K>1$, i.e., by zero drag at $T=0$. 
 The basin of attraction of these fixed points is the area below the separatrix $\lambar_s(K)=\sqrt{2}\left(\frac{1-K}{K}+\log{K}\right)^{1/2}$. Systems with bare couplings
$\lambar_0, K_0$ inside this area scale towards weaker backscattering
coupling $\lambar$ when temperature decreases. Below we show that no realistic set of interaction parameters falls under this category.

When the bare couplings are outside this region
(i.e. $K_0<1$ or $\lambar_0 > \lambar_s(K_0)$ )
renormalization to lower temperature
drives the system into the  strong coupling stable fixed point,
where $\lambar\rightarrow\infty$ and $K \rightarrow 0 $.
In this case  backscattering  becomes 
dominant at sufficiently low temperatures 
and freezes the phase $\phi_{c-}$ to a minimum position.
 Translated to the double-wire system this means that
the charges adjust their relative displacement $\sqrt{2}\phi_{c-} = 
\phi_+ - \phi_-$ in such a way that the $2k_F$ inter-wire potential is minimal,
i.e. the system forms two inter-locked CDWs.

Under these conditions the
drag is very strong, as  pointed out by Nazarov and Averin
\cite{nazarov_averin}. The system's resistivity to a flow of unequal
currents in the two wires becomes infinite, in the limit of zero
temperature and infinite length. We elaborate on this subject in
section (\ref{low_temperature}).  

It is instructive to express the 
condition for 
weak and strong coupling in terms of the bare interaction parameters
$g_i,\gbar_i$. For reasonably weak interaction, $K \lapprox 1.6$,
the separatrix $\lambar_s(K)$ is well approximated by $\lambar_s(K)
\approx  K-1 \approx U_{c-}$. Within this approximation the
requirement for strong coupling, $\lambar_0 > \lambar_s(K_0)$, states  
\ben\label{condition}
 \lambar_0 > U_{c-}^0.
\een
which is equivalent to 
\ben\label{strongly_coupled}
(g_2 - g_1) > ( \gbar_2 - \gbar_1) .
\een
Rather than the absolute strength of the 
inter-wire interaction couplings, it is their difference $\gbar_2 - \gbar_1$ in comparison to
$g_2-g_1$ which determines the zero temperature fixed point of the system. Taking $g_i$ and $\gbar_i$ as the Fourier components
of intra- and inter-wire interaction $V(x)$ and $\bar{V}(x)$ at $q_1 = 
2k_F$ and $q_2 = 1/L \to 0$, the condition \Ref{strongly_coupled} becomes
\ben
\int \! dx \: (1 - \cos q_1 x)\left[V(x) - \bar{V}(x)\right]
>0 .
\een
The first factor is non-negative. Then, a
sufficient condition for the l.h.s. to be positive is obviously 
$V(x) > \bar{V}(x)$ for all $x$, which is fulfilled by all
monotonously decaying repulsive interactions potentials, in particular 
the Coulomb potential. Therefore, {\it a Coulomb coupled double wire system
of spin-less electrons should scale towards strong coupling, which
implies a diverging zero temperature drag in infinitely long wires.}

There are two types of initial values $\lambar_0,K_0$ which flow to
the strong coupling fixed point. The first is defined by $K_0<1$. As
seen in Eqs. (\ref{flow}), for this case $\lambar$ varies monotonously
as the  temperature decreases.  The second type is defined by $K_0>1$
and $\lambar_0>\lambar_s(K_0)$. For this type, $\lambar$ does not varies
monotonously. For relatively high temperature, $\lambar$ decreases. At the
temperature at which $K=1$  $\lambar$ starts increasing
towards the strong coupling fixed point. As explained in the previous
subsection, well-separated wires ($\dbar\rightarrow\infty$) fall under
the first category, while wires at very close proximity fall under the
second.  
Assuming the bare interaction parameters $g_i,\gbar_i$ to be small,
for both types of initial conditions we may separate between a weak
coupling, high temperature, regime, where perturbation theory
calculations can be carried out, and a low temperature, strong
coupling regime. In the next subsections we calculate the drag
resistivity for both regimes, and identify the temperature scale that
separates the two.

\subsection{The high temperature regime}
\label{high_temperature}
We begin with the weakly coupled regime, in which we employ a method
devised by Nazarov and Averin \cite{nazarov_averin}. In the limit of
linear response ($I \to 0$) this approach  
of calculating the drag resistivity is similar to the memory-function 
formalism of Zheng and MacDonald \cite{zheng}, as it is shown in 
appendix \ref{equivalence}. The main difference is that in the present
calculation only the backscattering component of the inter-wire
interaction is treated perturbatively, while the small-momentum 
part is treated exactly.

We consider a
four-probe measurement with voltage probes at positions $x_0$ and
$x_0+a$  on both wires (let $u_{c-}/T \ll a \ll L$), and
calculate  the voltage drop $eV_w = 
\av{\mu_w(x_0)-\mu_w(x_0+a)}_I$ along wire $w$ when a current $I$ is
driven only through the active wire ($+$).
Using the relation $\delta eV_w = \kappa\inv \delta n_w$, where
 $\kappa\inv = \partial \mu_w / \partial n_w$ is the inverse 
compressibility, we obtain,
\be
eV_w = \fr{1}{\kappa} \av{ \delta n_w(x+a) - \delta n_w(x)}_I 
= - \fr{1}{ \kappa} \int_{x_0}^{x_0+a} \av{\partial_x n_w}_I.
\ee
Due to translational invariance, 
\ben\label{drag}
\fr{eV_w}{a} = \fr{1}{\kappa} \av{\partial_x n_w}_I
   =-\fr{\pi}{\sqrt{2} \kappa} \av{\dx^2 \phi_{c+} + w\dx^2 \phi_{c-}}_I.
\een
The thermo-dynamical averaging $\av{\dots}_I$ has to be restricted to
states 
satisfying $\av{I_+} = I$ and $\av{I_-}= 0$. Equivalently, but
technically more convenient, one can use an 
ensemble of current-less states, and then perform a Galilean-transformation
of the active wire such that a net current $I$ results. In terms of
the displacement fields $\phi_s$ this means that $\phi_+$ acquires a component growing
linearly in time, 
\be
\phi_+(x,t) \to \phi_+(x,t) + \Omega t,
\ee
or, translated into absolute and relative fields,
\ben\label{trans}
\phi_{c\pm} \to \phi_{c\pm} + \fr{\Omega t}{\sqrt{2}}.
\een
The frequency $\Omega$ is related to the current by $\Omega = \pi I/
e$ \cite{remark2}.

As expected, the transformation \Ref{trans} does not alter the absolute sector
$H_{c+}$, and no symmetric voltage is induced by the current. 
It does affect the relative Hamiltonian $H_{c-}$  via the
backscattering interaction, which becomes
\ben\label{H_int}
H_{int} = \lambar E_0 \int \fr{dx}{\pi \alpha} \:
\cos(\sqrt{8}\phi_{c-} + 2 \Omega t),
\een 
and gives rise to a finite drag-voltage
\ben \label{drag-voltage}
\fr{eV_w}{a} = w \fr{\pi\kappa}{\sqrt{2}}
\av{\dx^2\phi_{c-}}_{H_{int}}.
\een
For the following calculation it
is advantageous \cite{nazarov_averin} to make use of the
equation of motion, 
$\dt^2\phi_{c-} = -[H_{c-},[H_{c-}, \phi_{c-}]]$, from which
follows that
under stationary conditions ($\av{\dt^2\phi_{c-}} = 0$)
\bean
\av{\dx^2\phi_{c-}} &=& -\fr{K}{u} \av{ \fr{\delta H_{int}}{\delta \phi }}
 \label{dHdphi}\\
 &=& \fr{\sqrt{8} K}{\alpha^2}
\:\lambar \av{\sin( \sqrt{8} \phi + 2\Omega t )}, \nn
\eean
and then to perform a perturbative expansion of the r.h.s in the 
backscattering coupling $\lambar$ with respect to $H_0 =  H_{c-} -
H_{int}$. 

By standard methods we obtain in lowest non-vanishing order the 
resistivity
\ben\label{resistivity}
\rho_w   \equiv \left. \fr{\partial}{\partial I} \fr{ V_w}{
                l}\right|_{I=0} 
  = w \rho_0 \lambar^2 \left( \fr{T}{E_0} \right)^{4K - 3},
\een
where $ \rho_0 \sim \: h/e^2 \alpha$. Higher order terms
$\rho^{(n)} \propto  \lambar^{2n}$ scale with temperature as
\be
\rho^{(n)} \propto T^{\delta_n}, \quad \delta_n = (4K -4)n + 1.
\ee
Terms with odd powers of $\gbar_1$ vanish.

The temperature scaling law $\rho \propto T^{4K-3}$ 
can be also derived from the RG treatment in the following way:
In the absence of intra-wire interaction a simple calculation yields
\be
\rho(T) \sim \rho_0 \lambar_0^2 T/E_0.
\ee
The effect of intra-wire scattering can be accounted for  by using a
renormalized backscattering coupling constant $\lambar = \lambar(T)$.
One finds from the RG equations \Ref{flow} that in the weak coupling
limit $\lambar(T) = \lambar_0 \: (T/E_0)^{2K-2}$ (see appendix
\ref{tstar}, Eq. \Ref{lambda_of_x}),
which inserted into the previous equation indeed gives 
$\rho(T) \propto \lambar_0^2 \:(T/E_0)^{4K-3}$.

\subsection{The crossover temperature}\label{xovert}
In this section we estimate the cross-over temperature below which the perturbative calculation of the previous section ceases to hold, and inter-locking of the two charge density waves in the two wires becomes relevant. The cross-over temperature is that in which the coupling constant $\lambar$ becomes of order $1$. 

For initial conditions $\lambar_0\ll 1$ and $K_0<1$, corresponding to well separated wires, the solution to the RG equations can be approximated by $\lambar(T)\approx\lambar_0\left(\frac{E_0}{T}\right)^{2-2K_0}$ and $K\approx K_0$. The cross-over is then 
\ben
T^*\sim E_0\lambar_0^{1/(2-2K_0)}
\label{tsone}
\een

In appendix \ref{tstar} it is shown
that for $\dbar\gg d,k_F$ and for the Coulomb interaction,  the coupling constant $\lambar_0\sim \exp{-k_F\dbar}$, and thus $T^*$ is exponentially suppressed. 
More specifically, we find in this case
\ben\label{tstar_one}
T^* \sim E_0 \exp\left( -\fr {k_F \bar{d}}{1-K_0}\right),
\een
Note that with vanishing interaction strength, $1-K_{c-} \approx
-U_{c-} \approx r_s 
(\ln k_F \bar{d} + \gamma)\pi^{-2}$ (see section \ref{experiments}),
and hence $T^* \propto \exp(- \frac{k_F\dbar}{r_s\ln{k_F \dbar}})$,
where the parameter $r_s \equiv 
r / a_B = \pi e^2/ v_F \epsilon$ describes the
interaction strength.

For initial conditions of the type $K_0\gapprox 1$ and
$\lambar_0>\lambar_s(K)$, corresponding to wires at very close
proximity, the RG equations can again be solved approximately (see
Appendix (\ref{tstar}) for details), and we find,  
\ben\label{tstar_two}
T^* \sim E_0 \exp\left(-\fr{\pi^3}{r_s}\fr{c(k_F)}{k_F \dbar}\right),
\een
where $c(k_F)$ is of order one and only logarithmically depends on
$k_F$. Thus, when  $1/k_F \dbar \gg 1$, the temperature $T^*$ is again
exponentially  small. At first sight, this might be 
surprising. However, it becomes 
understandable by observing that for $\dbar \ll r$ intra and inter
wire coupling are almost equal. Hence the problem becomes close to that of 
a single wire with electrons of two spin states, (where $g_{i} =
\gbar_{i}$), which is known to renormalize
to the marginal weak coupling fixed point. 

\subsection{Low temperature limit}\label{low_temperature}
In 
the strong coupling limit,  with $\lambar \gapprox 1$,
and $K \ll 1$, the Sine-Gordon Hamiltonian \Ref{H_relative} describes
two inter-locked charge density waves, and carries strong resemblance
to the problem of  a pinned charge density wave, which has been 
studied  quite extensively in the 1970s (we refer in
particular to the work of Rice et al. \cite{rice} and Maki \cite{maki}).

Classically, for large $\lambar$ the Hamiltonian $H_{c-}$ has an infinite number of ground states
with field configurations 
$\phi(x) \equiv \phi_N = \pi(2N+1)/\sqrt{8}$. They are
separated by a large  energy barrier, such that at sufficiently low
temperatures the field fluctuates around
one of the ground states $\phi_N$. These fluctuations do not carry an anti-symmetric current. Rather, it is a transition from a ground
state $\phi_N$ to its  neighbor $\phi_{N\pm 1}$ which corresponds to an
electron transfer through the active wire from the left end to the right
end or vice versa. These transitions are carried out either by a soliton moving along the wire, or by a soliton-antisoliton pair which is created in the wire and is dissociated  by an electric field. The width $W$ and energy $E_s$ of a classical soliton are\cite{ramajaran}
\be
W \sim \fr{\alpha}{\sqrt{K\lambar}}, \quad E_s =
\sqrt{\fr{8 \lambar}{\pi^2 K}} E_0.
\ee
We expect that the transport properties of short wires with $L
\ll W$ are  different from that of long wires, where $L
\gg W$, and therefore consider both regimes separately.

In  long wires electronic current is carried by means of
soliton-antisoliton pairs. This mechanism was studied by Rice et
al. \cite{rice} and Maki \cite{maki} within a semi-classical
approximation. The semi-classical regime is that in which the
classical soliton energy $E_s$ is much larger than the zero point
energy of the fluctuations around it. In terms of our parameters this
regime is defined by $K\ll 1$.  
As found by Rice et al. \cite{rice}, in this regime thermal creation
of soliton-antisoliton pairs leads to a resistivity  
\ben\label{long_thermal_resistance}
\rho = \fr{h}{16\pi e^2l} \sqrt{\frac{E_s T}{2\lambar K E_0^2}} e^{E_s/T},
\een
where $l$, the mean free path of (anti)solitons, is a phenomenological
parameter. 

When $K\ll 1$, the soliton energy is renormalized 
according to \cite{tsvelik}
\ben
E_s^{r}=E_s\left(\frac{E_s}{E_0}\right)^{\frac{K}{1-K}}.
\label{renormsol}
\een
Eqs. \Ref{long_thermal_resistance} and \Ref{renormsol} define a
crossover temperature $T^*\equiv E_s^{r}$ separating the low and high
temperature regimes. As one may expect, this temperature is
approximately equal to the one obtained in Subsec. \ref{low_temperature}.

The low temperature regime is characterized 
by a highly non linear drag, taking place at exceedingly low
temperatures, where soliton-antisoliton pairs are generated by quantum
tunneling, rather than thermal activation. In this regime, as found by
Maki,  
\ben\label{long_tunneling_resistance}
\rho = \fr{h}{e^2} \fr{ E_s}{16\pi^2 u_{c-}} e^{\epsilon_0/\epsilon},
\een
where $\epsilon_0 = E_s^2/2 e u_{c-}, \epsilon = V/L$, and $V$ is the voltage difference between the two ends of the wires. The cross-over
temperature, at which thermally activated behavior changes to a
tunneling dominated one, 
can be determined by equating the exponentials, which yields
\be
T_{co} \sim 2 \sqrt{\fr{K}{\lambar}} \fr{\alpha}{L} eV.
\ee

For short wires, where
the soliton width $W$ is large compared to the wire length $L$, the
fields $\Pi,\phi$ may be approximated as constant in space, leading to
an Hamiltonian  
\bean
H &=& \frac{u_{c-}L}{2\pi}\left[{\pi^2
    K_{c-}}\Pi^2+\frac{2\lambar}{\alpha^2}\cos{\sqrt{8}\phi_{c-}}\right ] 
    \nn\\ 
& \equiv & \fr{\omega^2}{E_b} p^2 + \fr{E_b}{2} \cos \varphi
\label{pcjj}.
\eean
Here we introduced a frequency $ \omega^2 = 8 \lambar K E_0^2 $,
a barrier energy $E_b = \fr{2}{\pi}\lambar \fr{L}{\alpha}E_0$, and
canonical conjugated variables $\varphi = \sqrt{8} \phi$, $p = L \Pi /
\sqrt{8}$.
While this Hamiltonian resembles that of a pendulum, it
differs from it in one important aspect: in the present case the
states $\varphi$ and $\varphi \pm 2\pi$ are distinct physical states, since a
shift of the phase by $2\pi$ corresponds to a transfer of one electron
between the  reservoirs the active wire is coupled to. Since different
minima of the potential energy correspond to distinct states of the
reservoirs, it is unlikely that the field $\phi$ can be in a
superposition of several such minima. We therefore assume that the
field is in one of the minima, and transport takes place by tunneling
or thermal hopping between adjacent minima.  

A chemical potential difference $\delta\mu$ between
the reservoirs at the ends of the active wire adds a linear potential to the
Hamiltonian \Ref{pcjj} 
\be
V(q) = \fr{\delta\mu}{2\pi} \varphi .
\ee
To determine the current induced by $\delta\mu$ we calculate the rates
$\Gamma_{l/r}$ by which the phase $\varphi$ hops between minima by
thermal activation or tunneling. Then 
\be
I = e (\Gamma_r - \Gamma_l).
\ee

Thermal activated hopping dominates at temperatures $T \gg \omega$
with a rate \cite{kleinert}
\be
\Gamma_{l/r} \approx \fr{\omega}{2\pi} e^{-E_b/T} e^{\mp \delta\mu/ 2T}.
\ee
At low temperatures $T \ll \omega$ the main contribution comes again from 
tunneling processes with rate
\be
\Gamma'_{l/r} \sim \omega e^{-S_{l/r}(\delta \mu)}.
\ee
For $\delta \mu \ll E_b$ the action can be expressed as
\bea
S_{l/r}(\delta \mu) &=& \fr{E_b}{\sqrt{2} \omega} 
     \int_\sigma^{2\pi -\sigma} d\varphi  
\sqrt{1-\cos \varphi \: \pm \fr{\delta \mu }{\pi E_b}} \\
&\approx & \fr{4E_b}{\omega} \pm C \fr{\delta \mu}{\omega},
\eea
where $\sigma \sim (K \alpha^2 /\lambar L^2)^{1/4} \ll 1$ takes care
of the finite width of the well ground state. The coefficient $C$ is
of order unity and only logarithmically dependent on $\sigma$.

Putting this together we end up with current-voltage relations
\bean
I & \approx &  e  \fr{\omega}{\pi} e^{-E_b/T} \sinh( \delta\mu/ 2T ), \quad
\mbox{for } \quad T \gg \omega \label{short_thermal_resistance} \\
I & \sim &  e \omega e^{ -4E_b/\omega  } \sinh( C \delta\mu/ \omega), \quad
\mbox{for}\quad T \ll \omega. 
\label{short_tunneling_resistance}
\eean
Accordingly, the linear resistance is
\bean
R & \approx & \fr{h}{e^2} \fr{T}{\omega} e^{E_b/T}, \quad \mbox{for} \quad 
T \gg \omega, \label{short_thermal_R}  \\
R & \sim & \fr{h}{e^2} e^{4E_b/\omega}, \quad \mbox{for} \quad T \ll \omega.
\label{short_tunnel_R}
\eean
These expressions are 
valid for wires much shorter than the soliton length $W \sim \alpha/
\sqrt{K\lambar} $ and voltages $eV \ll E_b$. Notice that because 
of $E_b \propto L/\alpha$ the resistance increases exponentially with 
length $L$.

\section{Spin-full double wire}\label{spinfull}
\subsection{The Hamiltonian and renormalization group equations}\label{sfham}
We now extend our analysis to include the spin degree of freedom of electrons in the two wires.
Denoting the spin degree of freedom by an index $s=\pm$ in addition 
to $r,w$ as above, and taking into account the $SU(2)$ symmetry of the Coulomb interaction, we have 
\bea
H_0 &=& \fr{\pi v_F}{L} \sum_{qrws} \rho_{rsw}(q)\rho_{rsw}(-q), \\ 
H_f &=& \fr{1}{L} \sum_{qsws'w'} 
[\delta_{w,w'} g_2 +\delta_{w,-w'} \gbar_2] \rho_{+sw}(q)\rho_{-s'w'}(-q) \\
& + & \fr{1}{2L} \sum_{rqsws'w'} 
[\delta_{w,w'} g_4 +\delta_{w,-w'} \gbar_4]\rho_{rsw}(q)\rho_{rs'w'}(-q),\\
H_b &=& \sum_{sws'w'} \int dx\:
[\delta_{w,w'} g_1 +\delta_{w,-w'} \gbar_1] \times \\
& &\qquad\qquad \psi_{+sw}\dag(x) \psi_{-s'w'}\dag(x) \psi_{+s'w'}(x)
\psi_{-sw}(x). 
\eea

Following the same procedure as in the spin-less case, we first 
change to bosonic field variables \cite{schulz}
\bea
\phi_{sw}(x) &=& -\fr{i\pi }{L} \sum_{q\neq 0} \fr{e^{-iqx-\alpha
    |q|/2  }}{q} ( \rho_{+sw}(q)  + \rho_{-sw}(q)),\\
\Pi_{sw}(x) &=& \fr{1}{L} \sum_{q\neq 0} e^{-iqx-\alpha |q|/2} (
\rho_{+sw}(q) - \rho_{-sw}(q)),
\eea
and then transform  to new bosonic fields $\tilde\phi_{c/s\pm}, \tilde
\Pi_{c/s\pm} $ corresponding to absolute (+) and relative (-) charge
($c$) and spin ($s$) density. The transformation is defined by
\be
\phi_{sw} = \half(\tilde \phi_{c+} + s \tilde \phi_{s+} + w \tilde
\phi_{c-} + sw \tilde \phi_{s-} ),
\ee
and in the same way for $\tilde  \Pi_{c/s\pm}$.
For the transformation of the backscattering Hamiltonian $H_b$ we use
the formula
\ben\label{spin-fermions}
\psi_{rsw} = \fr{\eta_{rsw}}{\sqrt{2\pi \alpha}} \exp\left[ir(k_F x -
\phi_{sw}) + i\theta_{sw}\right], 
\een
where $\theta_{sw}(x_0) = \pi\inv \int_0^{x_0} \Pi_{sw} dx$ and the
$\eta_{rsw} = \eta_{rsw}\dag$ are so-called Majorana fermions. They
obey 
\be
[\eta_{rsw}, \eta_{r's'w'}]_+ = 2 \delta_{rsw,r's'w'}
\ee
and ensure that the operators given by \Ref{spin-fermions} 
follow fermionic commutation rules \cite{schulz}. We obtain the
following Hamiltonian (the 
tilde is omitted, $j$ takes the values $j=c,s$, for charge and spin. 
$l$ takes the values $l = \pm$, for symmetric and anti-symmetric)
\bean
H &=&  \sum_{jl} H_{jl} + H_b, \\
H_{jl} &=& \fr{u_{jl}}{2\pi} \int K_{jl} \pi^2 \Pi^2_{jl} +
              \fr{1}{K_{jl}} (\dx \phi_{jl})^2,\\ 
& & K_{jl} = (1+U_{jl})^{1/2}(1-U_{jl})^{-1/2} ,\\
& & \quad U_{c+} = \fr{1}{2\pi v_{c+}}( -2g_2 -2\gbar_2 + g_1), \\
& & \quad U_{c-} = \fr{1}{2\pi v_{c-}}( -2g_2 +2\gbar_2 + g_1), \\
& & \quad U_{s\pm} = \fr{1}{2\pi v_{s\pm}} g_1, \\
& & \quad \quad v_{c\pm} = v_F + (g_4 \pm \gbar_4)/\pi, \\
& & \quad \quad v_{s\pm} = v_F,\\
& & u_{jl} = v_{jl} (1-U_{jl}^2)^{1/2},\\
H_b & = & + \fr{g_1}{\pi^2 \alpha^2} \int dx\: \cos 2\phi_{s+} \:\cos
2\phi_{s-}\\
& & - \fr{\gbar_1}{\pi^2 \alpha^2} \int dx\: \cos 2\phi_{c-} \:\cos
2\phi_{s-} \\
& & - \fr{\gbar_1}{\pi^2 \alpha^2} \int dx\: \cos 2\phi_{c-} \:\cos
2\phi_{s+}.
\eean
Again, the system has no resistance to a symmetric flow of current, due to Galilean invariance, and thus the charge symmetric part of the Hamiltonian $H_{c+}$ decouples from the other parts.  
The remaining parts are all coupled by the backscattering Hamiltonian
$H_b$, so that the spin modes, despite their charge neutrality, do affect the role of the Coulomb interaction. 

As in the spin polarized case,  when the backscattering couplings $g_1,
\gbar_1$ scale to the strong coupling fixed point at $T=0$, spin
fields $\phi_{s\pm}$ and the relative charge field $\phi_{c-}$ freeze 
to minimum positions, which causes a divergent drag resistance
at zero temperature. On the other hand, when $\lambar$ renormalizes to zero
(while $\lambda$ can still scale to stronger couplings), the drag
resistance decreases with temperature and vanishes at $T=0$. 
(The case of a decreasing $\lambda$ and an increasing $\lambar$ does 
not exist.)

The second order RG-equations for a double wire as described by the
Hamiltonian above are
\bean
\fr{d \lambar}{d x} &=&  ( 2 - K_s - K_c - 2\lambda) \lambar, 
  \label{RGs_equation1} \\
\fr{d \lambda}{d x} &=& (2-2K_s)\lambda - 2\lambar^2,
  \label{RGs_equation2} \\
\fr{d K_s}{d x} &=& -2(\lambar^2 + \lambda^2)K_s^2,
  \label{RGs_equation3} \\
\fr{d K_c}{d x} &=& -4 \lambar^2 K_c^2,
  \label{RGs_equation4}
\eean
where $\lambar \equiv -\gbar_1/2\pi u_{c-}$, $\lambda \equiv
g_1/2\pi u_{s\pm}$, $K_c \equiv K_{c-}$, and $K_s \equiv K_{s\pm}$.
These equations can be derived in the same manner as those in the spin polarized case
by assuming scale invariance of the correlation functions 
$\av{ e^{2i\phi_{jl}(x_1,\tau_1)}e^{-2i\phi_{jl}(x_2, \tau_2)}}$
\cite{jose,giamarchi_schulz}. They are valid for arbitrary $K_{s/c}$,
but restricted to small $|\lambar|, |\lambda| \ll 1$.
(Similar RG equations appear in the context of single wall carbon
nanotubes, where also two spin-degenerate channels are present
\cite{egger_gogolin_yoshioka}.)

\subsection{Fixed points for the renormalization group equations}\label{sffp}
\begin{figure}[htb] 
 \begin{center}
         \epsfxsize 8cm
         \epsffile{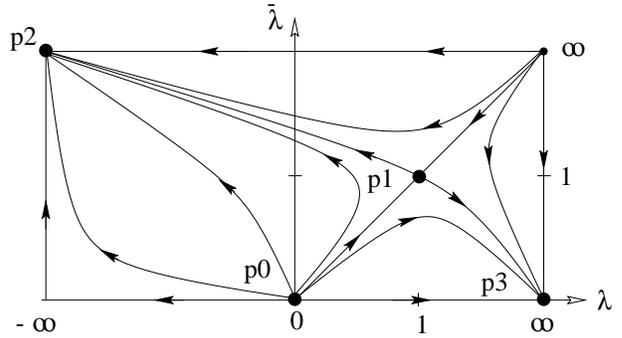}
 \end{center}
 \caption{ RG flow of the spin-full double wire system
   in the $K_c=K_s =0 $ plane (schematic). $p_0$ and $p_1$ are 
   unstable fixed points, $p_2$ is the strong coupling fixed point,
   and $p_3$ a spin-gap fixed point.
 }
 \label{fig-lambdaflow}
 \end{figure}

The following types of fixed points are found for the RG equations
\Ref{RGs_equation1}--\Ref{RGs_equation4}: 
\begin{itemize}
\item Fixed points where $K_s\ne 0$. As evident from
  \Ref{RGs_equation3}, for these fixed points $\lambda=\lambar=0$, and
  thus we refer to them as the "weak coupling fixed points". They
  describe wires with vanishing backscattering interaction at low
  temperatures, and therefore vanishing drag. As in the spin polarized
  case, for large  values of $K_c,K_s$ these fixed points are stable,
  while for small values they are unstable.  

\item Fixed points for which $K_s=0$ and $K_c\ne 0$. As evident from
  \Ref{RGs_equation4}, for these points $\lambar=0$, and then, from
  \Ref{RGs_equation2}, a stable fixed point exists only for
  $|\lambda|=\infty$. The former, $\lambar=0$, indicates vanishing
  drag, the latter, $|\lambda| = \infty$, means that the spin-modes
  are massive (spin gap). These fixed points  have no
  analog in the spin-less case.
  We refer to them as the  "spin-gap fixed points".   

\item Fixed points where $K_s=K_c=0$. There are two stable fixed points 
in this plane. The first is $\lambda=-\infty$ and $\lambar=+\infty$,
and it obviously describes two strongly coupled wires, with diverging
zero temperature drag. We refer to it as the "strong coupling fixed
point". The second is a spin-gap fixed point as described above. There
are also two unstable fixed points on that plane. The first is
$\lambda=\lambar=0$. It is repulsive in all directions of the
$(\lambda,\lambar)$ plane. The second is $\lambda=\lambar=1$. It is
attractive in the direction $(1,1)$. The border between the basin of
attraction of the two stable fixed points is the diagonal
$\lambda=\lambar$ (see Fig. \ref{fig-lambdaflow}). 
This border has a simple physical interpretation,
separating between the region where bare intra-wire backscattering is
stronger than the bare inter-wire one and the region where the
opposite is true.  
\end{itemize}
 
\subsection{The small $U$ limit}\label{sfsmallu}
In case of small $U_{c/s}$ Eqs. \Ref{RGs_equation1} to
\Ref{RGs_equation4} reduce to the following set of equations, derived
first  by  Gorkov and Dzyaloshinskii for the problem of coupled
chains\cite{gorkov,lee_rice_klemm}, 
\bean
\fr{d \lambar}{d x} &=&  -(U_s + U_c + 2\lambda) \lambar, \nn\\
\fr{d \lambda}{d x} &=& -2U_s\lambda - 2\lambar^2, \nn \\
\fr{d U_s}{d x} &=& -2(\lambar^2 + \lambda^2), \nn\\
\fr{d U_c}{d x} &=& -4 \lambar^2,\label{approx_RGs_equations}
\eean
($U_{c/s} \equiv U_{c-/s\pm}$).
Furthermore, in that limit $\lambda_0 = U_s^0$,
and $d( \lambda - U_s)/d x \equiv 0$. Therefore 
 $\lambda(x) = U_s(x)$.
With this, Eqs. \Ref{approx_RGs_equations} reduce
to three independent equations:
\bean
\fr{d \lambar}{d x} &=&  -(U_c +  3\lambda) \lambar, \nn\\
\fr{d \lambda}{d x} &=& -2(\lambda^2 + \lambar^2), \nn\\
\fr{d U_c}{d x} &=& -4 \lambar^2. \label{three_equations}
\eean
These RG flow equations were analyzed in detail by Lee et
al. \cite{lee_rice_klemm}, who found that for initial
values satisfying
\ben\label{strong_coupling_s}
U_c \lambda < \lambar^2,
\een
the system scales to a strong coupling fixed point $\lambda \to
-\infty$, $\lambar \to \infty$. Rewritten in terms of
$g_{i\dots},\gbar_{i\dots}$ this condition states
\be
\gbar_2 - g_2 < \half(\fr{\gbar_1^2}{g_1} - g_1) + O(g_{i\dots}^2/v_F^2)
\ee
For $g_1 > 0$ the r.h.s. is always larger than $\gbar_1 -
g_1$, such that a sufficient condition for this inequality to be fulfilled is
$\gbar_2 - g_2 < \gbar_1 - g_1$. This, however, is exactly the
inequality \Ref{strongly_coupled} we had in the spin-less case. We find then again that for weak  monotonously decaying repulsive electron--electron interaction (small positive $\lambda$),  
a double wire with spin unpolarized electrons scales towards strong
coupling, manifested by diverging trans-resistance at zero
temperature. The condition $\lambda_0=U_s^0$, valid in the small $U$
regime, does not allow a flow to the spin-gap fixed point.

\subsection{Beyond the small $U$ limit -- linear stability of the weak
  coupling fixed points}\label{sfallu} 
  
As stated above, a small $U$ analysis (i.e., weak electron--electron
interaction) of the RG equations
\Ref{RGs_equation1}--\Ref{RGs_equation4} leads to the conclusion that
$\lambar$ and therefore also the drag resistivity diverges at zero
temperature, accompanied by a large {\em negative} coupling
$\lambda$. It is unlikely that  
stronger electron--electron interaction would push the system towards
the weak coupling fixed points, where both backscattering interactions
become negligible at zero temperature. It is, however, conceivable
that for some range of parameters strong interaction would lead the
system towards the spin-gap fixed point, where $\lambda$ goes to 
large {\em positive} coupling, but $\lambar$ (and $\rho_D$) {\em
vanishes}. For example, if in the 
absence of inter-wire coupling each of the wires is in a spin--gapped
state with $\lambar \gapprox 1$ 
(as happens for attractive intra-wire electron--electron interaction) 
one may expect weak inter-wire coupling to leave the two wires in that
state. Consequently, the wires would decouple ($\lambar \to 0$) and
the drag would {\em vanish}. 

In principle one may analyze this behavior by use of the
RG equations \Ref{RGs_equation1}--\Ref{RGs_equation4} 
{\em beyond} the small $U$ limit. The validity of such an analysis to
the problem at hand is however unclear, since at the level of initial values
$\lambda_0>U_s^0$, and the RG equations are derived under the
assumption of small $\lambda$. Instead of doing this, we 
treat $K_{c/s}, \lambda$ and $\lambar$ 
as {\em independent} parameters and  confine ourselves to the case of
small couplings $\lambar, \lambda \ll 1$ (but arbitrary $K_{c/s}$),
where the RG equations are valid.

\begin{figure}[htb] 
 \begin{center}
         \epsfxsize 8cm
         \epsffile{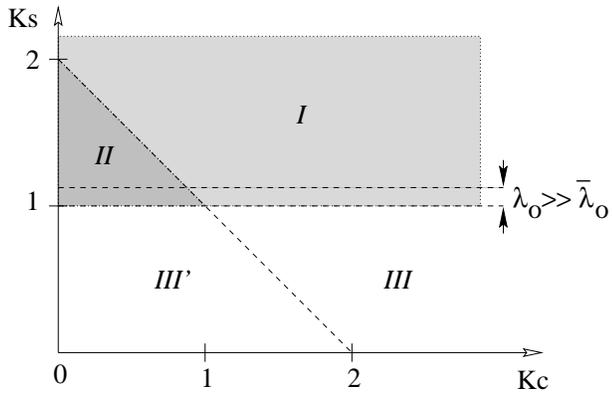}
 \end{center}
 \caption{ Stability of weak coupling fixed points $\lambda=\lambar=0$.
           Only fixed points in area $I$ can be stable. In the limit
           $|\lambar_0| \ll \lambda_0 \ll |K_s-1|$ points in area $II$
           flow to the strong coupling fixed point $p_2$ (where $\lambar 
           \to \infty$) , while points in area $III$ or $III'$ flow
           towards the spin-gap fixed point $p_3$ (where $\lambar \to 0$).
 }
 \label{fig-areas}
 \end{figure}

Let us consider the linear stability of the
weak coupling fixed points (where $\lambda=\lambar=0$) to turning on a
small $\lambda,\lambar$. We find that the
$\lambda=\lambar=0$ plane is split into four regions (see
Fig. \ref{fig-areas}).  In the first region, area $I$, defined by $K_s
+ K_c > 2$ and $K_s > 1$, the weak coupling fixed point is stable.  
In the region where $K_s>1$ but $K_s+K_c<2$ (which we call area $II$,
see Fig. \ref{fig-areas}), the weak coupling fixed points are  linearly  
stable with respect to infinitesimal values of $\lambda$, but unstable
with respect to such values of $\lambar$. We now argue that initial
values in that area flow to the strong coupling fixed point ($\lambda
\to -\infty, \lambar \to +\infty$). Let us
consider $|\lambar_0| \ll \lambda_0 \ll 
K_s-1$ in this region. By Eq. \Ref{RGs_equation2}, the coupling
constant $\lambda$ will scale to small values, such that the
instability of the $\lambar$-mode (Eq. \ref{RGs_equation1}) becomes
relevant . Then $\lambar$ keeps increasing until finally 
the $-2\lambar^2$-term in Eq. \Ref{RGs_equation2}
overcomes the $\lambda$-stability and forces $\lambda$ to scale to
negative values. 

Points lying below the 
$K_s=1$ line in area $III$ or $III'$ are linearly unstable in
$\lambda$. Consequently, $\lambda$ increases and thereby decreases the
coefficient in the equation for $\lambar$. In the limit of $\lambar_0
\ll \lambda_0$ this leads to a stabilization of the
$\lambar$-mode, even where it was initially unstable (area $III'$).
Hence, points in areas $III$ and $III'$ scale to
the spin-gap fixed point ($\lambda \to +\infty, \lambar =0$).

To summarize, in the limit of $|\lambar| \ll \lambda \ll |K_s-1|$ the
basins of attraction of the weak and strong coupling fixed points
correspond to areas I and II in Fig \ref{fig-areas},
respectively. Further, points of area $III/III'$, which characterize
systems with tendency to (single-wire) spin gap phase, indeed scale
towards the (double-wire) spin-gap fixed point with vanishing
$\lambar$ (zero drag).   

The linear stability analysis we carried out is expected to give the
right structure of Fig. \Ref{fig-areas}, but not the precise borders
between the areas. Near the borders terms linear in $\lambda$ and
$\lambar$ may have very small prefactors, which make the role of the
quadratic terms significant. An example to that is the region
$K_s\approx 1+\lambda$, studied by Lee {\it et al.}, where in the
right hand side of \Ref{RGs_equation2} both terms become comparable. 

When $\lambda_0$  is comparable to  $|K_s-1|$ (but still $\lambar_0 \ll
1$) the condition for $\lambda$ scaling towards weaker couplings is
$K_s - 1 > \lambda_0$ (see the discussion in Sect. \ref{spinless}).
Then, the
border between areas $III/III'$ and $II/I$ is raised
to $K_s = 1 + \lambda_0$. The effect of larger $\lambda_0$ is
therefore a reduction of area $II$, i.e. also in this sense
a larger $\lambda_0$ is in favor of wire decoupling.

\subsection{High and low temperature limits of the drag}\label{sfhlt}
A perturbative calculation of the drag-voltage in the weak coupling 
regime, applicable for widely separated wires in the high temperature
regime, can be developed along the same lines as before. We have  
\be
eV_w = \fr{a \kappa}{\pi} \av{\dx^2 \phi_{c+} + w \dx^2\phi_{c-}}_I.
\ee
The transformation to current carrying states is again given by
\Ref{trans}, such that
\bea
eV_w &=& w\fr{a\kappa}{\pi} \av{\dx^2\phi_{c-}}_I \\
&=& w \fr{4 a \kappa}{\pi \alpha^2} \lambar \av{\sin(2\phi_{c-} +
\sqrt{2} \Omega t)(\cos 2\phi_{s-} + \cos 2\phi_{s+})}.
\eea
Hereby, we used the Eq. of motion for $\phi_{c-}$ and the stationarity 
condition $\av{\dt^2\phi_{c-}} =0$. A perturbative
expansion in the couplings $\lambar$, $\lambda$ and $U_s \approx
\lambda$  then leads to 
\ben\label{rho-scaling}
\rho_w = w \rho_0 \:\lambar^2 \left( \fr{T}{E_0}
\right)^{2K_c - 1},
\een
with  $\rho \sim h/e^2 \alpha$. Leading higher order terms are of order
$ \lambar^2 \times  O( \lambda, U_s )$.

For wire distances $\dbar \gg k_F\inv$ the smallness of 
$\lambar \propto e^{-2k_F \dbar}$ is guaranteed. Further, in this case
the spin-sector couplings $\lambda, U_s$ flow to the single wire
Luttinger Liquid fixed point $\lambda^* = U_s^* = 0$, as long as the
temperature is still large compared to $T^*$. Hence, also effective 
couplings $\lambda \approx U_s$ are small, wherefore the perturbative
result \Ref{rho-scaling} is applicable.

The cross-over temperature $T^*$, separating between the high and low
temperature regimes, is exponentially suppressed for $\bar{d} \gg
k_F\inv$ with
\ben\label{tstar_three}
T^* \sim E_0 \exp \left( -\fr{2k_F \bar{d}}{ 1 -K_c } \right),
\een
as shown in appendix \ref{tstar}.

When the system flows towards the strong coupling fixed points, 
at sufficiently low temperatures the phases $\phi_{s\pm},\phi_{c-}$
freeze to their minimum positions,
 and anti-symmetric current flows by
means of solitons. Again, in this case we expect drag resistivity to
be proportional to $e^{T*/T}$.  

\section{Estimate of parameters}\label{experiments}
The significance of Coulomb drag in a particular experimental set-up of
two 1D coupled wires is determined by $T^*$ and  
the corresponding minimal length $L^*= \alpha E_0/T^*$. We now estimate
these quantities for a double wire system with 
 parameters taken from the experiment by Yacoby et al.
\cite{yacobi}, where, using cleaved edge over-growth in GaAs-AlGaAs structures, 
quantum wires of width down to $14nm$ and length
$L$ of order $\mu m$ were fabricated, with  adjustable electron density.

The bare values of the interaction constants
$g_{i\dots},\gbar_{i\dots}$ appearing in the fermionic Hamiltonians
can be estimated from the geometry of the experiment. 
If we assume the wires to be parallel in a distance $\dbar$ and
characterize the transversal extension of the electron wavefunction 
by a length $d$, intra- and inter-wire potential are given by
\be
V(x) = \fr{e^2}{\epsilon}(x^2  + d^2)^{-1/2},\quad 
\bar V(x) = \fr{e^2}{\epsilon}(x^2  + d^2 + \dbar^2)^{-1/2}.
\ee
The expression for $V(x)$ is approximate on small lengthscales
$\lapprox d$. However, since generally $k_F\inv \gapprox d$, this may
not lead to large errors in the determination of
the $2k_F$ scattering parameters, $g_1,\gbar_1$.
The couplings $g_i,\gbar_i$, are the Fourier
components of $V(x), \bar{V}(x)$ at $q_1=2k_F$ and $q_2, q_4 \to 0$,  
\be
g_i =  \fr{2e^2}{\epsilon} \Knull(q_i d),\quad \gbar_i =
       \fr{2e^2}{\epsilon} \Knull(q_i \sqrt{d^2+\dbar^2}).
\ee 
Taking the limit $q_1 \to 0$ does not raise any problem, since the
$q\to 0$ couplings 
$g_{2/4},\gbar_{2/4}$ appear always as differences. Screening by
metallic  gates used in the experiment is negligible due to the
relative large distance. 
As pointed out by Starykh {\it et al.} (\cite{starykh}, see also Mahan
\cite{mahan}), the parameter $g_4$ has to be modified for the
Pauli principle to be satisfied. This modification is however small
when the $2k_F$ part of the interaction is much smaller than the
$q\approx 0$ part.  

Having determined $g_{i\dots},\gbar_{i\dots}$, we can
calculate the bare parameters $\lambda_0, K_{c-}^0$, and
$\lambar_0, K_{s}^0$ for the bosonic Hamiltonians by means of the
corresponding expressions of Sect. \ref{spinless} and \ref{spinfull}. These
values are then to be used as initial values $(x=0)$ for the RG
equations.  

The system enters the regime of strong
coupling when $\lambda(x) \sim 1$ (spin-polarized) or
$\lambar(x) \sim 1$ (spin-unpolarized). Thus, $T^*$ and  $L^*$ 
are estimated by $\lambda(x^*)=1$ or $\lambar(x^*)=1$, respectively.
The integration of the RG-equations can
be done analytically in certain parameter regimes, which leads to the
expressions \Ref{tstar_one}, 
\Ref{tstar_two}, and \Ref{tstar_three} (see appendix \ref{tstar}).
Here we also give numerical results.

One should be aware that this method leads to order of magnitude
estimates rather than precise values for $T^*$. Nevertheless, the reliability
of this procedure can be demonstrated in the special 
case of spin-polarized fermions with $K_{c-} = 1/2$.
Then, the bosonic Hamiltonian $H_{c-}$
can be transformed to a Hamiltonian of non-interacting, fictitious
fermions (``refermionization'') \cite{luther_emery}, which 
exhibit an energy gap of the order of $T^*$ obtained by
the aforementioned method (see appendix \ref{refermionization}).

Let us first consider the spin-polarized case. 
We assume an inter-wire distance of $\bar{d} = 3d$.
Table \ref{table-one} contains $\lg_{10} E_0/T^* \equiv \lg_{10} L^*/r$
for wire widths $d = a_B, 2 a_B$ 
and $4 a_B$, where $a_B$ is the effective Bohr radius. With $a_B
\approx 10nm$ in GaAs these wire widths are close to the experimental
values in \cite{yacobi}. The TL-parameter of the relative charge
sector is typically $K_{c-} \approx 0.8$ and $\lambda_0 \lapprox 0.1$.
The main characteristic is the strong decrease of $x^*$ with $r$
as long $r \lapprox \dbar $, in accordance with the 
exponential suppression of $T^*$ (see Eq. \Ref{tstar_one}). For larger
values of $r$ the dependence is  
rather weak, since here, according to \Ref{tstar_two}, $x^* \approx
 \pi^3 c(k_F)/ r_s k_F \dbar $ with $c(k_F)/r_s k_F \approx const.$.

The strong coupling regime is not easily accessible. Even at very low
densities of $r\inv = (200nm)\inv$ the crossover temperature
is still very low: $T*\sim 0.01 E_0 \sim 1mK$ (the
corresponding length is $L^* \sim 10\mu m$). However, even when
the system is still in the weak coupling regime, the drag resistance 
can be significant. According to the Eq. \Ref{resistivity} we estimate
the drag resistance of a $10\mu$ long double wire at $T=100mK$
with $d=10nm$, $\dbar=3d$, and $r/a_B = 5,10$ and 20 to be of order
0.01, 0.1 and 1$\times h/e^2$, respectively.

Quantitatively, our findings deviate strongly from the estimate for
$L^*$ given by Nazarov and Averin \cite{nazarov_averin}. For parameters 
$d=\bar d = 10nm$, $D=100nm$ Ref.  \cite{nazarov_averin} obtain $L^*\approx 0.3\mu m$,
whereas according to our calculations for these parameters $L^*\approx
1000\mu m$. The origin of this discrepancy of more than three orders
of magnitude is explained in the next section.

The results obtained for a double wire of spin-unpolarized electrons is
shown in Tab. \ref{table-two} (still $\dbar= 3d$, $D=200nm$).
For relatively small values $r\lapprox \dbar$ the system scales to
strong inter wire couplings $\lambar$. In this regime, $x^*$
decreases with increasing 
$r$, as in the previous case. A qualitatively different behavior sets
in at larger $r$: here, renormalization drives the system towards the 
spin gap fixed point, indicated  
by a ``$*$'' in Tab. \ref{table-two}. The transition from the former
regime to the latter happens at $r_c/a_B =3.1, 4.8$, and 7.6 for
$d/a_B = 1.0, 2.0$, and 4.0, respectively. At these densities, 
the TL parameter assume values $K_{c-} \approx 0.5$ and $K_s \approx
1.3$. Since further $\lambda \approx 0.3$, it is unclear, however, to what
extent the RG equations we use, which are derived for small
$\lambda_0,\lambar_0$ are applicable for this case.  

\section{Summary}\label{comparison}
In this paper we used the Tomanaga--Luttinger model to analyze Coulomb
drag between two ballistic quantum wires. We find the drag to be a
strong effect, both in its magnitude and in its temperature
dependence.  

We find that at zero temperature, for all weak  monotonously decaying
repulsive electron--electron interaction the trans-resistance diverges,
indicating the formation of inter-locked charge density wave ground
state in the two wires. At low temperature, $T\ll T^*$, we predict the
trans-resistance to depend exponentially on $T^*/T$.  

At high temperature $T \gg T^*$ we predict the trans-resistance to show
a power law dependence on temperature, with the power being determined
by the Luttinger liquid parameters. For spin polarized electrons, we
find the power to be $4K_{c-}-3$, with $K_{c-}$ being the TL-parameter
corresponding to {\it anti-symmetric}  charge 
displacement. In this, our findings are in contrast to that of
Ref. \cite{nazarov_averin} where the exponent was identified with the
parameter corresponding to {\it symmetric} charge
displacement.
 Quantitatively, the two parameters are different,
and this difference leads to the big difference in the estimates
for $T^*$. For spin unpolarized electrons, we find the power to be
$2K_{c-} -1 $.

The crossover temperature $T^*$ depends exponentially on
parameters. For wires at large separation $k_F\dbar\gg 1$ it is
$T^*\sim E_0\exp{-\frac{b k_F\dbar}{1-K_{c-}^0}}$, where $b=1$
for spin-unpolarized, and $b=2$ for spin-polarized wires.

We thank I. Aleiner, A. Finkel'stein, D. Maslov,
A. Punnoose and Y. Oreg for  instructive discussions, and particularly
A. Moustakis, L. Balents and S. Simon for a discussion of the RG
equations. We thank the 
Minerva-Foundation (Munich), the Bi-national Israel--USA foundation,
the Israel academy of Science and the Victor Ehrlich chair for
financial support.  

\begin{appendix}

\section{}\label{tstar}
We begin with the calculation of $T^*$ for spin-less electrons and 
large wire separation $\dbar$, such that $k_F \dbar \gg 1$. In this
case
\be
\lambda_0 = \fr{2 e^2}{\pi \epsilon u_{c-}} \Knull(2 k_F \dbar)
\sim r_s e^{-2 k_F \dbar}
\ee
is exponentially small, as a consequence of which $K_{c-}(x) \approx
const. \equiv K_0$ (Eq. \ref{flow}). With this we obtain from 
 $\d \ln \lambda /\d x = 2 - 2K_0$
\ben\label{lambda_of_x}
\lambda(x) = \lambda_0 e^{(2-2K_0)x},
\een
where $x = \ln E_0/T$. The condition $\lambda(x^*) \approx 1$ then gives
\ben\label{rg_gap}
T^* \approx E_0 \lambda_0^{1/(2-2K_0)} \sim E_0
\exp\left(-\fr{k_F\bar{d}}{1-K_0} \right).
\een

For small wire separations ($d\approx \dbar$) inter and intra wire
couplings $\gbar_{\dots} $ and $g_{\dots}$ become similar, such that 
($U_0\equiv U_{c-}^0$)
\be
U_0 = \fr{1}{2 \pi v_{c-}} ( g_1 - g_2 + \gbar_2 ) \approx 
\fr{g_1}{2 \pi v_{c-}} \approx \fr{\gbar_1}{2\pi u_{c-}} = \lambda_0.
\ee
This means that the bare couplings $\lambda_0, K_0 \approx 1+U_0$  lie
close to the separatrix, but still outside the attractive region  
of the weak coupling fixed points (see Fig. \ref{fig-flow}). Under
renormalization $(\lambda(x), K(x))$ flows along the separatrix
towards weaker couplings until the turning point at $K=1$ is
reached. Thenceforward $(\lambda, K)$ flows towards the strong
coupling regime. 

For an estimate of $T^*$ we use the approximate RG equations 
valid for $|U| \ll 1$,
\be
\fr{\d \lambda}{\d x} = -2 \lambda U, \quad \fr{\d U}{\d x} = -2
\lambda^2.
\ee
These can be easily integrated by use of the constant of motion 
$ \lambda^2 - U^2 \equiv a^2$, which yields 
\be
x_2 - x_1  = \fr{1}{2a}( \arctan U_1/a - \arctan U_2/a ).
\ee
In particular, we obtain for the ``time'' $x_a$, at which the turning
point is reached ($x_1 = x_0 =0, U_1 =U_0, x_a = x_2, U_2 =0$)
\be
x_a = \fr{1}{2a} \arctan U_0/a.
\ee
In the case under consideration with $\lambda_0 \approx U_0$, the
constant $a$ is a small quantity:
\be
a^2 \approx 2\lambda_0(\lambda_0 - U_0) \ll \lambda_0^2, U_0^2,
\ee
which will be determined more explicitly later. Because of the
relative smallness of $a$ we have
\be
x_a \approx \fr{\pi}{4a}.
\ee
For the rise from the turning point to stronger couplings $\lambda
\gg a$, an equally long ``time'' $x_b$ is needed, so that
\be
x^* \approx  x_a + x_b = \pi/2a
\ee
and therefore
\ben\label{tstar_a}
T^* \approx E_0 e^{-\pi/2a}.
\een
It remains to calculate $a$: Taking the expressions
\be
g_i = \fr{2 e^2}{\epsilon} \Knull(q_i d),\quad 
\gbar_i = \fr{2 e^2}{\epsilon} \Knull(q_i \dbar),
\ee
and expanding up to second order in $k_F \dbar$ we obtain
\bea
\lambda_0 - U_0 &=& \fr{1}{2\pi v_{c-}} (\gbar_1 - g_1 + g_2 -
\gbar_2) \\
&=& \fr{r_s}{\pi^2} f(d/\dbar, \ln k_F \dbar)\:(k_F \dbar)^2  +
O^4(k_F \dbar),
\eea
whereby
\be
f(d/\dbar, \ln k_F \dbar) 
= ( \fr{d}{\dbar} \ln d/\dbar + \gamma -1 +\ln k_F
\dbar)(\fr{d^2}{\dbar^2}-1).
\ee
($\gamma \approx 0.577$ is Euler's constant)
This yields
\be
a = \left(2 \lambda_0 ( \lambda_0 - U_0)\right)^{1/2} =
\fr{r_s}{\pi^2} \Knull^{1/2}(2k_F\dbar) f^{1/2}\: k_F \dbar.
\ee
As long $k_F \dbar \lapprox 1$, the Bessel-function $\Knull(2k_F
\dbar)$ and $f$ are of order one and only logarithmically
dependent on $k_F \dbar$. Hence in this case
\be
a = c\inv \fr{r_s}{2\pi^2} k_F \dbar
\ee
with $c$ a numerical coefficient of order unity. Inserting this into
\Ref{tstar_a} yields
\be
T^* \sim E_0 \exp \left( - \fr{c \pi^3}{r_s k_F \dbar} \right).
\ee

Let us now consider a spin-full double wire with large separation
$\dbar \gg k_F\inv$. Then again the inter wire backscattering coupling 
is exponentially small, such that also here
\be
K_c \approx const. = K_c^0.
\ee
Due to the smallness of $\lambar$, the RG equations for the spin-sector
couplings $\lambda, K_s$ decouple from the relative charge sector and
become identical to the corresponding RG equations of a single
spin-$\half$ wire:
\bea
\lambda' &=& -2 U_s \lambda, \\
U_s' &=& - 2 \lambda^2
\eea
(for $\lambda, |U_s| \ll 1$).
Accordingly, the couplings $\lambda, U_s$ with bare values $\lambda_0 =
U_s^0 = g_1/2\pi v_F$ scale down to weaker coupling:
\be
\lambda(x) = U_s(x) = \fr{\lambda_0}{1+2\lambda_0 x}.
\ee
Because of this behavior we can neglect in the differential equation
for $\lambar$,
\be
\fr{\d \ln \lambar}{\d x} = 1 - K_c^0 - U_s - 2 \lambda,
\ee
the last two terms, if $U_s^0, \lambda_0 \ll 1 - K_c^0s$. 
Then, for this regime, 
\be
\fr{\d \ln \lambar}{\d x} \approx 1 - K_c^0,
\ee
which gives
\be
\lambar(x) = \lambar_0 e^{(1-K_c^0)x}.
\ee
This leads in the same way as in the corresponding spin-less case 
to a crossover temperature
\be
T^* \sim E_0 \exp\left( - \fr{2k_F \bar{d}}{1-K_c^0}\right).
\ee
This result deviates from the former estimation for the spin-less case
only by a factor 2 in front of $k_F \bar{d}$. This extra factor 
reflects the fact that here the mean electron distance $\bar{r}$ is
exactly half of the one in a spin-less wire, when the Fermi momentum is
the same.

\section{Relation to the Memory-Function Formalism}\label{equivalence}
In this section we elaborate on the relation 
of the perturbative calculation of subsection \ref{high_temperature}
and a formula for the Coulomb drag resistivity based on the
memory-function formalism by Zheng and MacDonald (Eq. (12) in \cite{zheng}).

From Eqs. \Ref{drag-voltage} and \Ref{dHdphi} follows that 
\be
 \rho = M \fr{ \partial }{\partial \Omega} \av{ \fr{ \delta
     H_{int}(\Omega) }{ \delta \phi(x_0)} },
\ee
where $M$ is a constant factor determined by the system parameters, 
and $H_{int}$ is given by Eq. \Ref{H_int}.
A first order expansion in $H_{int}=H_{int}(\Omega)$  leads
to  
\be
\av{ \fr{ \delta  H_{int} }{ \delta \phi(x_0)}} \approx i \int_0^\infty dt
 \av{[ H_{int}(t) , \fr{ \delta H_{int}(t=0) }{ \delta \phi(x_0)} ]}_0.
\ee
where the subscript $0$ indicates a thermal average  taken with respect to $H_0$. 
For $I \propto \Omega \to 0$ we can expand $H_{int}$ in $\Omega$,
\be
H_{int}(\Omega) 
 \approx H_{int}(\Omega=0) + \fr{\Omega t}{\sqrt{2}} \int \fr{ \delta H_{int}}{
  \delta \phi(x)} dx,
\ee
and obtain
\be
 \rho = i\fr{M}{\sqrt{2}} \int_0^\infty dt \  t\ \int dx \av{
   [\fr{ \delta H_{int}(t) }{ \delta \phi(x)},
   \fr{ \delta H_{int}(0) }{ \delta \phi(x_0)}]}_0.
\ee 
Then, making use of the Kubo--identity 
\bea
[ e^{-\beta H_0}, A] &=& i e^{-\beta H_0} 
\int_0^\beta d \lambda  \fr{ d A }{dt}(-i\lambda) \\
&=& i e^{-\beta H_0} \int_0^\beta d \lambda  e^{\lambda H_0}
    \fr{ dA }{dt}e^{-\lambda H_0}
\eea
and partial integration leads to
\be
\rho =  \fr{M}{\sqrt{2}} \int_0^\infty dt \int_0^\beta d\lambda \int dx
 \av{
     \fr{ \delta H_{int}(-i\lambda)}{ \delta \phi(x_0)}
     \fr{ \delta H_{int}(t) }{ \delta \phi(x) }
    }_0.
\ee
Since the functional derivatives of $H_{int}$ are the 
intra-wire force-densities (up to a constant factor), this expression
essentially equals the corresponding formula (12) of Zheng and
MacDonald. 

\section{Refermionization}\label{refermionization}
Luther and Emery \cite{luther_emery} observed that in the special case
$K_{c-} = 1/2$ the bosonic Hamiltonian $H_{c-}$ can be mapped  onto an
exactly solvable Hamiltonian of fictitious, {\em noninteracting} fermions
(``refermionization''). 
After the canonical transformation $\Pi_{c-}, \phi_{c-} \to \Pi' =
2^{1/2} \Pi_{c-}, \phi' = 2^{-1/2}\phi_{c-}$ the Hamiltonian states 
\be
H_{c-} = \fr{u_{c-}}{2\pi} \int dx\: \pi^2 \Pi'^2 + (\dx \phi')^2 
+ \fr{2\gbar_1}{2\pi \alpha} \int dx\: \cos(2 \phi'),
\ee
which expressed by fermions 
\bea
a_r(k) &=& 2^{-1/2} \int dx\: \psi_r(x) e^{-ikx} \\
\psi_r &=& (2\pi \alpha)^{-1/2} \exp( ir(k_F x + \phi') + i \pi \int^x_{x_0}
d x' \Pi'(x')),
\eea
becomes
\bea 
H_{c-}^f &=& u_{c-} \sum_{rk} rk a_r\dag(k) a_r(k) \\
& & + \fr{\gbar_1}{2\pi\alpha} \sum_k  a_+\dag(k) a_-(k-2k_F) +h.c. .
\eea
This Hamiltonian is diagonziable by a simple Bogoliubov
transformation. The spectrum 
\be
E_{\pm,k} = u_{c-} k_F \pm \sqrt{ \left( \fr{\gbar_1}{2\pi \alpha}
\right)^2 + u_{c-} (k -k_F)^2}
\ee
exhibits an energy gap of width $\Delta E = g/\pi \alpha$ at $\pm
k_F$, which coincides up to a factor 2 with the corresponding
temperature $T^*$ as estimated by integrating the RG flow:
from Eq. \Ref{rg_gap} with $K_{c-}= 1/2$ we get
\be
T^* = E_0 \lambda_0 = E_0 \fr{\gbar_1}{2\pi u_{c-}} = \fr{\gbar_1}{2\pi 
\alpha} = \half \Delta E.
\ee

\end{appendix}

\begin{table}[h] 
\be
\begin{array}{c|ccc}
 d/a_B & 1.0 & 2.0 & 4.0 \\ \hline
 r/a_B & & & \\
 2.0    & 15  &  27 & -   \\
 3.0    & 8.2 & 14  & -   \\
 4.0    & 5.6 & 9.1 & 17  \\
 5.0    & 4.3 & 6.7 & 11  \\
 7.0    & 3.1 & 4.4 & 7.4 \\
 9.0    & 2.6 & 3.3 & 5.3 \\
 12.   & 2.2 & 2.6 & 3.8 \\
 16.   & 2.0 & 2.1 & 2.8 \\
 20.   & 1.8 & 1.9 & 2.3 \\
    \end{array}
\ee
 \caption{\label{table-one}
  The table lists  $\lg_{10} E_0/T^*$  (which is also $\lg_{10}
  L^*/r$) for different values of $d/a_B$ and $r/a_B$ for a
  spin-polarized  
  double wire system ($\dbar = 3d$, $D=200nm$, $a_B = 10nm$). }
\end{table}

\begin{table}[h]
\be
\begin{array}{l|ccc}
 d/a_B & 1.0 & 2.0 & 4.0 \\ \hline
 r/a_B &     &     &    \\
 1.0   & 18  & -   & -  \\
 1.5   & 11  & 18  & -  \\
 2.0   & 8.7 & 13  & 22 \\
 2.5   & 7.6 & 9.9 & 17 \\
 3.0   & 7.3 & 8.4 & 13 \\
 3.5   & *   & 7.6 & 11 \\
 4.0   & *   & 7.0 & 9.7 \\ 
 4.5   & *   & 6.7 & 8.7 \\
 5.0   & *   & *   & 8.0 \\
 5.5   & *   & *   & 7.5 \\
 6.0   & *   & *   & 7.1 \\
 6.5   & *   & *   & 6.8 \\
 7.0   & *   & *   & 6.6 \\
 7.5   & *   & *   & 6.8 \\
 8.0   & *   & *   & *  
    \end{array}
\ee
 \caption{ \label{table-two}
  $\lg_{10} E_0/ T^* \equiv \lg_{10} L^*/\alpha$ for a spin-unpolarized   
  double wire system ($\dbar = 3d$, $D=200nm$, $a_B = 10nm$). The
  ``$*$'' indicates the zero-drag phase at $T=0$. Note that at large
  $r/a_B$ the initial values of $\lambda,\lambar$ are not small, and
  thus the validity of the RG equations is questionable. }  
\end{table}

\newpage

\end{multicols}

\begin{thebibliography}{10}

\def \pra#1#2#3#4{ #1, Phys.~Rev.~A {\bf #2}, #3 (#4)}
\def \prb#1#2#3#4{ #1, Phys.~Rev.~B {\bf #2}, #3 (#4)}
\def \prl#1#2#3#4{ #1, Phys.~Rev.~Lett. {\bf #2}, #3 (#4)}

\bibitem{2Ddrag} For a review, see  A. G. Rojo, J. Phys. Condens
  Matter, 11, R31-R52 (1999), and references therein. 

\bibitem{tomanaga__luttinger} S. Tomanaga, Prog. Theor. Phys. {\bf 5}
,544 (1950); J. M. Luttinger, J. Math. Phys. {\bf 4}, 1154 (1963).

\bibitem{haldane} F. D. M. Haldane, J. Phys. C {\bf 14}, 2585 (1981);
\prl{F. D. M. Haldane}{47}{1840}{1981}.

\bibitem{voit} J. Voit, Rep. Prog. Phys {\bf 57}, 977 (1994).

\bibitem{emery} V.J. Emery, {\em Highly Conducting One-Dimensional
  Solids } ed. J.T. Devreese, R.E. Evrard and V.E. von Doren, New York,
  Plenum (1979).

\bibitem{solyom} J. Solyom, Adv. Phys. {\bf 28}, 201 (1978).

\bibitem{schulz} H. J. Schulz, G. Cuniberti, P. Pieri,
  {\em Fermi liquids and Luttinger liquids}
  Lecture notes of the Chia Laguna summer school, Italy (1997). 

\bibitem{hu_flensberg} Ben Yu-Kuang Hu and Karsten Flensberg, in {\em
  Hot Carriers in Semiconductors}, edited by K. Hess et al., Plenum
  Press, New York, 1996.

\bibitem{raichev} \prl{O.E. Raichev, P. Vasilopoulos}{83}{3697}{1999}.

\bibitem{tanatar} \prb{B. Tanatar}{58}{1154}{1998}.

\bibitem{flensberg} \prl{K. Flensberg}{81}{184}{1998}.

\bibitem{komnik} \prl{A. Komnik, R. Egger}{80}{2881}{1998}.

\bibitem{nazarov_averin} \prl{Yuli V. Nazarov, D. V. Averin}{81}{653}{1998}.

\bibitem{lee_rice_klemm} \prb{P.A. Lee, T.M. Rice, R.A. Klemm}{15}{2984}{1977}.

\bibitem{double_chain}
  \prb{A.M. Finkel'stein, A.I. Larkin}{16}{10461}{1993};
  \prb{H.J. Schulz}{53}{R2959}{1996};
  \prb{L. Balents, M.P.A. Fisher}{53}{12133}{1996}; and references therein.

\bibitem{luther_emery} \prl{A. Luther and V. J. Emery}{33}{589}{1974}.

\bibitem{chui_lee} \prl{S. T. Chui and P. A. Lee}{35}{315}{1975}.

\bibitem{rice} \prl{M.J. Rice, A.R. Bishop, J.A. Krumhansl,
   S.E. Trullinger}{36}{432}{1976}.

\bibitem{maki} \prl{K. Maki}{39}{46}{1977}.

\bibitem{tarucha} S. Tarucha et al., Sol. State Comm. {\bf 94}, 413 (1995).

\bibitem{yacobi} \prl{A. Yacoby et al.}{77}{4612}{1996}.

\bibitem{jose} \prb{J.V. Jose, L.P. Kadanoff, S. Kirkpatrick,
    D.R. Nelson}{16}{1217}{1977}.

\bibitem{giamarchi_schulz} \prb{T. Giamarchi, H.J. Schulz}{37}{325}{1988}.

\bibitem{remark2}
  In principle, also the conjugate field $\Pi_+$ changes, nameley
  to $\Pi_+(x,t) + \fr{\Omega}{\pi v_F}. $
  However, $\Pi_+$ appears to the second power in the kinetic
  energy term. Therefore the transformation of $\Pi_+$
  contributes to the drag only into second order in $I$ resp. $\Omega$.
  We will neglect this
  contribution from the very beginning by leaving $\Pi_+$ unchanged.

\bibitem{zheng} \prb{Lian Zheng and A. H. MacDonald}{48}{8203}{1993}.

\bibitem{ramajaran}  R. Rajaraman, {\em Solitons and Instantons},
  North-Holland Publishing Company, Amsterdam (1982).

\bibitem{tsvelik} A. Tsvelik, {\em Quantum field Theory in Condensed Matter
Physics}, Cambridge University Press (1995).

\bibitem{kleinert} H. Kleinert, {\em Pathintegrals in Quantum
Mechanics, Statistics, and Polymer Physics}, 2nd Edition, World
Scientific Publishing, Singapore (1994).

\bibitem{egger_gogolin_yoshioka} R. Egger, A.O. Gogolin,
      Phys. Rev. Lett. 79, 5082 (1997), and Eur. Phys. J. {\bf B 3},
      281 (1998); 
      \prl{H. Yoshioka,A. A. Odintsov}{82}{374}{1999}.

\bibitem{gorkov} L.P. Gor'kov, I.E. Dzyaloshinskii,
Zh. Eksp. Teor. Fiz. {\bf 67}, 397 (1974) [Sov. Phys.-JETP {\bf 40},
198 (1975)].
\bibitem{starykh}  O. A. Starykh, D. L. Maslov, W. H\"ausler and
  L. I. Glazman cond-mat/9911286, to appear in "Interactions and
  Transport Properties of Lower Dimensional Systems", 
     Lecture Notes in Physics, Springer.
\bibitem{mahan} G.D. Mahan, {\it Many particle physics}, 2nd edition,
  Plenum press, New-York, 1990, section 4.4 .

\end{thebibliography}
\end{document}